\documentclass[english]{article}
\usepackage[T1]{fontenc}
\usepackage[latin9]{inputenc}
\usepackage{geometry}
\usepackage{bm}
\usepackage{amsmath}
\usepackage{graphicx}
\usepackage{esint}
\usepackage[numbers]{natbib}

\makeatletter
\newcommand{\lyxaddress}[1]{
	\par {\raggedright #1
	\vspace{1.4em}
	\noindent\par}
}

\usepackage{lineno}

\makeatother

\usepackage{babel}
\begin{document}
\title{Theoretical research on low-frequency drift Alfvén waves in general
Tokamak equilibria}
\author{Yang Li$^{1,2}$}
\maketitle

\lyxaddress{1. Southwestern Institute of Physics, PO Box 432, Chengdu 610041,
People\textquoteright s Republic of China}

\lyxaddress{2. Consorzio CREATE. Via Claudio, 21 80125 Napoli, Italy}

leeyang\_fusion@qq.com
\begin{abstract}
We developed kinetic models based on general fishbone-like dispersion
relations. Firstly, a general model for arbitrary magnetic configuration
and ion orbit width is presented. Then, by disregarding ion orbit
width and approximating the magnetic geometry as circular, we introduce
a simplified model that fully incorporates circulating/trapped ion
effects. Finally, by considering the limit of ions being well-circulating
or deeply trapped, the results directly revert to those observed in
earlier theoretical studies.
\end{abstract}

\section{Introduction}

Alfvén waves and energetic particles, resulted from fusion reaction
and auxiliary heating, are crucial to the performance of Tokamak devices.
The theoretical research on low frequency drift Alfvén waves (DAW)
is based on the general fishbone-like dispersion relation (GFLDR)
\citep{chen_physics_2016,zonca_theory_2014,zonca_theory_2014-1} and
gyrokinetic theory\citep{brizard_foundations_2007,frieman_nonlinear_1982}.
Besides recovering diverse limits of the kinetic magnetohydrodynamic
(MHD) energy principle, the GFLDR approach is also applicable to electromagnetic
fluctuations, which exhibit a wide spectrum of spatial and temporal
scales consistent with gyrokinetic descriptions of both the core and
supra-thermal plasma components. Formally ,the GFLDR can be written
as 
\begin{equation}
i\Lambda=\delta W_{f}+\delta W_{k},
\end{equation}
where $\Lambda$ is the generalized inertia, and $\delta W_{f}$ and
$\delta W_{k}$ are, respectively, fluid and kinetic potential energy
of electromagnetic fluctuations. In previous researches, $\Lambda$
is calculated with fluid\citep{chen_energetic-particle_2017} and
kinetic approaches\citep{chavdarovski_effects_2009}. Furthermore,
analytical findings regarding mode frequency, damping, and the impacts
of finite orbit width can be corroborated through numerical verification,
demonstrating a high level of agreement for both SAW continuum structure
\citep{bierwage_gyrokinetic_2017,lauber_analytical_2018,choi_gyrokinetic_2021,falessi_shear_2019}
and and geodesic acoustic mode (GAM) oscillations\citep{gao_eigenmode_2008,biancalani_cross-code_2017}.
As a result, the GFLDR, serving as a useful theoretical framework,
aids in comprehending experimental observations, numerical simulations,
and analytical outcomes across varying degrees of approximation.

In the original theoretical works\citep{zonca_theory_1996,zonca_existence_1999}
on DAW, ions considered in the kinetic analysis are assumed to be
well circulating. Later on, the kinetic analysis was extended by including
the deeply trapped ions and electrons\citep{chavdarovski_effects_2009,chavdarovski_analytic_2014}.
Moreover, the researches mentioned above are all based on the $s-\alpha$
model in Tokamak plasmas with circular configuration. However, the
effects of general magnetic geometry and full circulating/trapped
particles are not included in previous researches. Especially, the
particles near circulating/trapped separatrix are not included in
the previous theoretical models. In order to obtain a better understanding
of experimental observations\citep{heidbrink_stability_2021,gorelenkov_energetic_2014}
and provide a more precise kinetic model for theoretical researches\citep{ma_low-frequency_2023},
we need to include the general magnetic geometry and full orbit effects
without assuming well-circulating/deeply-trapped ions and small ion
orbit width. 

In this work, we present a kinetic model with general magnetic geometry
and arbitrary ion orbit width. With this general model, one can solve
the problem of DAW in the inertial layer numerically as long as the
guiding center orbit information is provided. By taking small ion
orbit width limit and applying the $s-\alpha$ model, a simplified
mode with full circulating/trapped ions effects is obtained. The generalized
inertia with the modification of neoclassical effects can also be
calculated with appropriate circular geometry data. Furthermore, if
we assume ions are well circulating or deeply trapped, the results
go back to those of the previous researches\citep{chavdarovski_effects_2009}. 

The rest of the paper is organized as follows. In Section 2, we present
our general kinetic model including guiding center motion in general
geometry, kinetic equation in ballooning space and governing equations.
Section 3 is devoted to solve the governing equation of the general
model inertial layer. A simplified model including the full circulating/trapped
ions effects is presented in Section 4 for circular magnetic geometry
by assuming small ion orbit width. Finally, the results are summarized
and discussed in Section 5.

\section{Theoretical model}

\subsection{Guiding center motion}

The guiding center motion of single particles can be given as
\begin{equation}
\dot{\mathbf{X}}=v_{\parallel}\mathbf{b}-v_{\parallel}\mathbf{b}\times\nabla\left(\frac{v_{\parallel}}{\Omega_{s}}\right),
\end{equation}
where $v_{\parallel}$ is the velocity parallel to magnetic field,
$\Omega_{s}=e_{s}B/m_{s}c$ is the gyro-frequency of the s-species
and $\bm{b}$ is the unit vector of magnetic field. In general model,
the magnetic field configuration of 2-D asymmetric equilibrium can
be denoted by the straight field aligned flux $\left(\psi,\theta,\zeta\right)$,
where $\psi$ is the poloidal flux, $\theta$ is the poloidal angle,
$\zeta=\phi-\nu\left(\psi,\theta\right)$, $\phi$ is the geometric
toroidal angle and $\nu$ is a periodic function in $\theta$. Thus,
the magnetic field is given as 
\begin{equation}
\bm{B}_{0}=F\left(\psi\right)\nabla\phi+\nabla\phi\times\nabla\psi.
\end{equation}
 The safety factor can be defined as $q\left(\psi\right)=\bm{B}_{0}\cdot\nabla\zeta/\bm{B}_{0}\cdot\nabla\theta$
and the Jacobian ${\cal J}=\left(\nabla\psi\times\nabla\theta\cdot\nabla\zeta\right)^{-1}$.
Then the particle orbit in the flux coordinates can be given as 
\begin{equation}
\dot{\psi}_{s}=\dot{\bm{X}_{s}}\cdot\nabla\psi=-\frac{v_{\parallel}}{{\cal J}B}\frac{\partial\bar{\psi}_{s}}{\partial\theta},
\end{equation}
\begin{equation}
\dot{\theta}_{s}=\dot{\bm{X}_{s}}\cdot\nabla\theta=\frac{v_{\parallel}}{{\cal J}B}\frac{\partial\bar{\psi}_{s}}{\partial\psi},
\end{equation}
\begin{equation}
\dot{\zeta_{s}}=\dot{\bm{X}_{s}}\cdot\nabla\zeta=\frac{qv_{\parallel}}{{\cal J}B}+\frac{v_{\parallel}m_{s}c}{{\cal J}Be_{s}}\left[\frac{\partial}{\partial\psi}\left({\cal J}B_{0}v_{\parallel}-q\frac{Fv_{\parallel}}{B}\right)+\frac{\partial}{\partial\theta}\left({\cal J}v_{\parallel}\frac{\nabla\psi\cdot\nabla\theta}{BR^{2}}-\frac{Fv_{\parallel}}{B}\frac{\partial\nu}{\partial\psi}\right)\right],
\end{equation}
where $R=\left|\nabla\phi\right|^{-1}$ and $\bar{\psi}_{s}=\psi-Fv_{\parallel}/\Omega_{s}$.
As we can see, $\bar{\psi}_{s}$ is a constant along the particle
orbit. In order to study the responses of trapped and circulating
particles, the canonical angle can be introduced as\citep{zonca_nonlinear_2015}
\begin{equation}
\vartheta_{c}=\omega_{bs}\int^{\vartheta_{0}}\frac{d\vartheta_{0}^{\prime}}{\dot{\theta}_{s}},\label{can-angle}
\end{equation}
where the bouncing frequency $\omega_{bs}=2\pi/\left(\ointclockwise d\vartheta_{0}/\dot{\theta_{s}}\right)$,
and $\ointclockwise$ means the integration along the particle orbit
with constant $\bar{\psi}_{s}$.

\subsection{Kinetic equation in ballooning space}

Before going to ballooning space, we start the analysis from the gyro-kinetic
equation in original space. The perturbed particle distribution function
can be written as\citep{frieman_nonlinear_1982,chen_physics_2016}
\begin{equation}
\delta f_{s}=\left(\frac{e}{m}\right)_{s}\left[\frac{\partial F_{0}}{\partial{\cal E}}\delta\phi-J_{0}\left(k_{\perp}\rho_{Ls}\right)\frac{QF_{0}}{\omega}\delta\psi e^{iL_{k}}\right]_{s}+\delta K_{s}e^{iL_{ks}},
\end{equation}
where $\delta\phi$ is perturbed electric potential, $\delta\psi$
is related to parallel vector potential fluctuation $\delta A_{\parallel}$
with $\delta A_{\parallel}=-i\left(\frac{c}{\omega}\right)\bm{b}\cdot\nabla\delta\psi$,
$k_{\perp}^{2}=k^{2}-\left(\bm{k}\cdot\bm{b}\right)^{2}$, $J_{0}$
is the 0th Bessel function, $\rho_{Ls}=m_{s}cv_{\perp}/e_{s}B_{0}$
the Lamor radius, $QF_{0s}=\left[\omega\partial_{{\cal E}}+\Omega_{s}^{-1}\left(\bm{k}\times\bm{b}\right)\cdot\nabla\right]F_{0s}$,
and $L_{ks}=\left(m_{s}c/e_{s}B_{0}\right)\left(\bm{k}\times\bm{b}\right)\cdot\bm{v}$.
And the non-adiabatic part $\delta K_{s}$ obeys the gyro-kinetic
equation as 
\begin{align}
\left(-i\omega+\dot{\psi}\partial_{\psi}+\dot{\theta}\partial_{\theta}+\dot{\zeta}\partial_{\zeta}\right) & \delta K_{s}=i\left(\frac{e}{m}\right)_{s}QF_{0s}\times\nonumber \\
 & \left[J_{0}\left(\delta\phi-\delta\psi\right)+\left(\frac{\omega_{d}}{\omega}\right)_{s}J_{0}\delta\psi+\frac{v_{\perp}}{k_{\perp}c}J_{1}\delta B_{\parallel}\right],\label{gyro-k}
\end{align}
where $J_{1}$ is the 1st Bessel function, $\omega_{ds}=\left(\bm{k}\times\bm{b}\right)\cdot\left(\mu\nabla B_{0}+v_{\parallel}^{2}\bm{\kappa}\right)/\Omega_{s}$,
$\bm{\kappa}$ is the curvature of magnetic field line and $\delta B_{\parallel}$
is the perturbed parallel magnetic field. By considering the ballooning
representation, we have\citep{chen_physics_2016} 
\begin{align}
f\left(\psi,\theta,\zeta\right) & =e^{in\zeta}\sum_{m}e^{-im\theta}f_{m,n}\nonumber \\
 & =e^{in\zeta}\sum_{m}e^{-im\theta}\int e^{-i\left(nq-m\right)\vartheta}\hat{f}d\vartheta.\label{balloon-rep}
\end{align}
Then the gyro-kinetic equation (\ref{gyro-k}) can be rewritten as
\begin{align}
\left[-i\omega+\dot{\theta}\partial_{\vartheta}+i\omega_{ds}\right] & \delta\hat{K}_{s}=i\left(\frac{e}{m}\right)_{s}QF_{0s}\nonumber \\
\times & \left[J_{0}\left(\delta\phi-\delta\psi\right)+\frac{\omega_{ds}}{\omega}J_{0}\delta\psi+\frac{v_{\perp}}{k_{\perp}c}J_{1}\delta B_{\parallel}\right],\label{balloon-k}
\end{align}
where $\omega_{ds}=-nq^{\prime}\dot{\psi}_{s}\vartheta+n\left(\dot{\zeta}_{s}-q\dot{\theta}_{s}\right)$
in ballooning representation. For the simplicity of notation, $\omega_{ds}$
denotes the drift frequency in ballooning space. Before further analysis,
it should be noted that the analysis in this work will be carried
out in the inertial layer. Thus, we have two scales, i.e. $\left|\vartheta_{1}\right|\sim\delta^{-1}\sim\beta^{-1/2}$
and $\left|\vartheta_{0}\right|\sim1$, where $\delta$ is the small
parameter. In the original space , $\dot{\theta}_{s}$, $\dot{\zeta}_{s}$
and $\dot{\psi}_{s}$ are all periodic functions in $\theta$. Since
periodic function in $\theta$ and be represented as periodic function
in $\vartheta$, all the periodic functions in original space, can
be treated as functions only in the short scale $\vartheta_{0}$,
which indicates that the particle orbit is on the scale of $\vartheta_{0}$. 

\subsection{Governing equations}

The governing equations for low frequency AE problem include mainly
vorticity equation and quasi-neutrality equation. Formally, the vorticity
equation can be given as\citep{zonca_resonant_2016}
\begin{equation}
\text{FLB}+\text{ICU}+\text{MPC}+\text{KPC}+\text{MFC}=0,\label{general-vor}
\end{equation}
where $\text{FLB}$ is finite line bending term, $\text{ICU}$ is
inertia-charge uncovering term, $\text{MPC}$ is MHD non-adiabatic
particle compression term. $\text{KPC}$ is kinetic particle compression
term, $\text{MFC}$ is magnetic field compression. And these terms
can be expressed as 
\begin{equation}
\text{FLB}\simeq\bm{B}\cdot\nabla\left(\frac{k_{\perp}^{2}}{k_{\vartheta}^{2}B^{2}}\bm{B}\cdot\nabla\delta\hat{\psi}\right)=\frac{1}{{\cal J}}\partial_{\vartheta}\left(\frac{k_{\perp}^{2}}{{\cal J}k_{\vartheta}^{2}B_{0}^{2}}\partial_{\vartheta}\delta\hat{\psi}\right),\label{flb}
\end{equation}
\begin{align}
\text{ICU} & \simeq-\frac{4\pi\omega^{2}}{k_{\vartheta}^{2}c^{2}}\sum_{s}\left\langle \frac{e_{s}^{2}}{m_{s}}\frac{QF_{0s}}{\omega}\left(1-J_{0}^{2}\right)\delta\hat{\phi}\right\rangle ,\label{icu}
\end{align}
\begin{equation}
\text{MPC}=\frac{4\pi}{k_{\vartheta}^{2}c^{2}}\sum_{s}\frac{e_{s}^{2}}{m_{s}}\left\langle J_{0}^{2}\omega\omega_{ds}\frac{QF_{0s}}{\omega}\delta\hat{\phi}\right\rangle _{v},\label{mpc}
\end{equation}
\begin{equation}
\text{KPC}=-\sum_{s}\frac{4\pi e_{s}}{k_{\vartheta}^{2}c^{2}}\left\langle J_{0}\omega\omega_{ds}\delta K_{s}\right\rangle _{v},\label{kpc}
\end{equation}
\begin{equation}
\text{MFC}=\frac{4\pi\omega^{2}}{k_{\vartheta}^{2}c^{2}}\sum_{s}\left\langle \frac{v_{\perp}}{k_{\perp}c}J_{0}J_{1}\frac{e_{s}^{2}}{m_{s}}\frac{QF_{0s}}{\omega}\delta\hat{B}_{\parallel}\right\rangle _{v},\label{mfc}
\end{equation}
where $\left\langle \cdots\right\rangle _{v}=2\pi\sum_{\sigma}\int\int B/\left|v_{\parallel}\right|\left(\cdots\right)d{\cal E}d\mu$
. Here we have assumed $F_{0s}$s are isotropic in pitch angle and
used the parallel Ampère's law 
\begin{equation}
\frac{k_{\perp}^{2}}{k_{\vartheta}^{2}}\bm{b}\cdot\nabla\delta\psi=\frac{4\pi}{k_{\vartheta}^{2}c^{2}}i\omega\sum_{s}e_{s}\left\langle v_{\parallel}\delta f_{s}\right\rangle _{v},\label{amp-law}
\end{equation}
and perpendicular Ampère's law
\begin{equation}
\delta B_{\parallel}\simeq\frac{4\pi}{B^{2}}\frac{c}{\omega}\left(\bm{k}\times\bm{b}\cdot\nabla P_{\perp}\right)\delta\psi,\label{amp-law-perp}
\end{equation}
where $P_{\perp}=P_{\perp i}+P_{\perp e}$ is the total perpendicular
pressure. By re-defining $\left[\delta\Psi,\delta\Phi,\delta\hat{B}_{\parallel}\right]=\frac{k_{\perp}}{k_{\vartheta}}\left[\delta\psi,\delta\phi,\delta B_{\parallel}\right]=\hat{\kappa}_{\perp}\left[\delta\psi,\delta\phi,\delta B_{\parallel}\right]$
and considering $\hat{\kappa}_{\perp}\sim s\left|\vartheta_{1}\right|\sim\delta^{-1}$,
the vorticity equation can be rewritten as 
\begin{equation}
\left(\frac{\partial^{2}}{\partial\vartheta^{2}}-\frac{\partial_{\vartheta}^{2}\left|\nabla\psi\right|}{\left|\nabla\psi\right|}\right)\delta\Psi+\frac{{\cal J}^{2}B^{2}}{\hat{\kappa}_{\perp}}\left(\text{ICU}+\text{MPC}+\text{MFC}+\text{KPC}\right)=0.\label{vor-reduced}
\end{equation}
Is should be noted that all the terms $\delta\psi$, $\delta\phi$
and $\delta B_{\parallel}$are substituted by $\delta\Psi$, $\delta\Phi$,
and $\delta\hat{B}_{\parallel}$ respectively in Eq. (\ref{vor-reduced}).
And the quasi-neutrality equation is 
\begin{equation}
\sum_{s}\left\langle \delta f_{s}\right\rangle _{v}=0.\label{quasi-neutr}
\end{equation}
By solving the governing equations (\ref{amp-law-perp})-(\ref{quasi-neutr}),
the physical property of DAW in the inertial layer can be determined.

\section{General solution in singular layer}

In the singular layer with $\vartheta_{1}\sim\delta^{-1}$, we have
the orderings $\delta\omega_{di}\sim\omega\sim\omega_{*Pi}\sim\omega_{bi}$
for circulating ions and $\omega_{de}\sim\omega_{di}\sim\omega\sim\omega_{bi}\sim\delta\omega_{be}$
for trapped ions and electrons, where $\omega_{*Pi}=\omega_{*Ti}+\omega_{*ni}$,
$\omega_{*Ti}=\left(T_{i}c/e_{i}B\right)\left(\bm{k}\times\bm{b}\right)\cdot\left(\nabla T_{i}\right)/T_{i}$,
$\omega_{*ni}=\left(T_{i}c/e_{i}B\right)\left(\bm{k}\times\bm{b}\right)\cdot\left(\nabla n_{i}\right)/n_{i}$,
$T_{i}$ is the ion temperature and $n_{i}$ is the ion density. Moreover,
the Lamor radius of ions is assumed to be small, i.e. $k_{\perp}\rho_{ti}\sim\delta$,
in singular layer. 

The 0th order kinetic equations for circulating ions, trapped ions
and trapped electrons can be written as
\begin{equation}
\left(\dot{\vartheta}\partial_{\vartheta_{0}}-i\omega\right)\delta K_{cir}^{\left(0\right)}=i\left(\frac{e}{m}QF_{0}\right)_{i}\frac{k_{\vartheta}}{k_{\perp}}\left(\delta\Phi^{\left(0\right)}-\delta\Psi^{\left(0\right)}\right),\label{0th-kic}
\end{equation}
\begin{equation}
\left(\dot{\vartheta}\partial_{\vartheta_{0}}-i\omega+i\omega_{di}\right)\delta K_{tr,i}^{\left(0\right)}=i\left(\frac{e}{m}QF_{0}\right)_{i}\frac{k_{\vartheta}}{k_{\perp}}\left(\delta\Phi^{\left(0\right)}-\delta\Psi^{\left(0\right)}+\frac{\omega_{di}}{\omega}\delta\Psi^{\left(0\right)}\right),\label{0th-kit}
\end{equation}
\begin{equation}
\left(-i\omega+i\omega_{de}\right)\delta K_{tr,e}^{\left(0\right)}=i\left(\frac{e}{m}QF_{0}\right)_{e}\frac{k_{\vartheta}}{k_{\perp}}\left(\delta\Phi^{\left(0\right)}-\delta\Psi^{\left(0\right)}+\frac{\bar{\omega}_{de}}{\omega}\delta\Psi^{\left(0\right)}\right),\label{0th-ket}
\end{equation}
where 
\begin{equation}
\bar{\omega}_{ds}=\frac{n\omega_{bs}}{2\pi}\ointclockwise\frac{\dot{\zeta_{s}}-q\dot{\theta_{s}}}{\dot{\theta_{s}}}d\vartheta_{0},\label{prec-freq}
\end{equation}
and $\delta K_{cir}^{\left(0\right)}$, $\delta K_{tr,i}^{(0)}$ and
$\delta K_{tr,e}^{\left(0\right)}$ are, respectively, 0th order non-adiabatic
kinetic responses for circulating ions, trapped ions and trapped electrons.
As shown in the equations above, the solutions for $\delta K_{circ}^{\left(0\right)}$
and $\delta K_{tr,e}^{\left(0\right)}$ can be readily written as
\begin{equation}
\delta K_{cir}^{\left(0\right)}=-\left(\frac{e}{m}\right)_{i}\frac{QF_{0i}}{\omega}\frac{k_{\vartheta}}{k_{\perp}}\left(\delta\Phi^{\left(0\right)}-\delta\Psi^{\left(0\right)}\right),\label{0th-kcir}
\end{equation}
and
\begin{equation}
\delta K_{tr,e}=-QF_{0e}\left(\frac{e}{m}\right)_{e}\frac{k_{\vartheta}/k_{\perp}}{\omega-\bar{\omega}_{de}}\left[\frac{\bar{\omega}_{de}}{\omega}\delta\Psi^{\left(0\right)}+\delta\Phi^{\left(0\right)}-\delta\Psi^{\left(0\right)}\right].\label{0th-kte}
\end{equation}
It should be noted that the solution (\ref{0th-kte}) is obtained
by neglecting the electron orbit width. By integrating over $\vartheta_{c}$
along particle orbit, the solution of Eq. (\ref{0th-kit}) can be
formally written as\citep{zonca_nonlinear_2015}
\begin{align}
\delta K_{tr,i}^{\left(0\right)}= & \frac{e^{-i\tilde{Q}_{b}}}{2\pi}\sum_{l}\frac{e^{il\vartheta_{c}}}{-i\omega+i\bar{\omega}_{di}+i\left(l-n\bar{q}\right)\omega_{bi}}\oint d\vartheta_{c}e^{i\tilde{Q}_{b}\left(\vartheta_{c}^{\prime}\right)-il\vartheta_{c}^{\prime}}\nonumber \\
 & \times i\left(\frac{e}{m}QF_{0}\right)_{i}\frac{k_{\vartheta}}{k_{\perp}}\left(\delta\Phi^{\left(0\right)}-\delta\Psi^{\left(0\right)}+\frac{\omega_{di}}{\omega}\delta\Psi^{\left(0\right)}\right),\label{0th-kti}
\end{align}
where 
\begin{equation}
\tilde{Q}_{b}\left(\vartheta_{c}\right)=\tilde{\Xi}\left(\vartheta_{c}\right)-n\tilde{\Pi}\left(\vartheta_{c}\right)-n\vartheta_{1}\left[q\left(\vartheta_{c}\right)-q\left(0\right)\right],
\end{equation}
\begin{equation}
\tilde{\Xi}=n\zeta\left(\vartheta_{c}\right)-\frac{\bar{\omega}_{di}}{\omega_{bi}}\vartheta_{c},
\end{equation}
\begin{equation}
\tilde{\Pi}=\int^{\vartheta_{0}\left(\vartheta_{c}\right)}qd\vartheta_{0}-\vartheta_{c}\bar{q},
\end{equation}
and $\bar{q}=\frac{1}{2\pi}\ointclockwise qd\vartheta_{0}$. As shown
in Eq. (\ref{0th-kti}), the resonance from toroidal and poloidal
orbit frequency can be included by $\omega_{di}$ and $\omega_{bi}$.
From Eq. (\ref{quasi-neutr}), the 0th order quasi-neutrality equation
can be given as
\begin{align}
\left(1+\frac{1}{\tau}\right)\left(\delta\Phi^{\left(0\right)}-\delta\Psi^{\left(0\right)}\right)= & \frac{T_{i}}{eN_{i}}\left\langle \frac{k_{\perp}}{k_{\vartheta}}\frac{1}{{\cal J}\omega_{bi}}\frac{\partial\bar{\psi_{i}}}{\partial\psi}\delta K_{circ}^{\left(0\right)}\right\rangle _{circ,\vartheta_{c}}\nonumber \\
-\frac{T_{i}}{eN_{i}}\left\langle \frac{k_{\perp}}{k_{\vartheta}}\frac{1}{{\cal J}\omega_{be}}\delta K_{tr,e}\right\rangle _{tr,\vartheta_{c}} & +\frac{T_{i}}{eN_{i}}\left\langle \frac{k_{\perp}}{k_{\vartheta}}\frac{1}{{\cal J}\omega_{bi}}\frac{\partial\bar{\psi_{i}}}{\partial\psi}\delta K_{tr}^{\left(0\right)}\right\rangle _{tr,\vartheta_{c}},\label{0th-quasi-neut}
\end{align}
where the averaged ion density $N_{i}=\left\langle \frac{F_{0i}}{{\cal J}\omega_{bi}}\frac{\partial\bar{\psi_{i}}}{\partial\psi}\right\rangle _{full,\vartheta_{c}}$,
$\left\langle \cdots\right\rangle _{full,\vartheta_{c}}=\left\langle \cdots\right\rangle _{circ,\vartheta_{c}}+\left\langle \cdots\right\rangle _{tr,\vartheta_{c}}$
$\left\langle \cdots\right\rangle _{circ,\vartheta_{c}}=\sum_{\sigma}\int\int\oint\left(\cdots\right)d\vartheta_{c}d{\cal E}d\mu$
within the circulating domain of $\left({\cal E},\mu\right)$, $\left\langle \cdots\right\rangle _{tr,\vartheta_{c}}=\int\int\oint\left(\cdots\right)d\vartheta_{c}d{\cal E}d\mu$
the integration within the domain of trapped particles\citep{hinton_theory_1976},
$\sigma=v_{\parallel}/\left|v_{\parallel}\right|$ and the summation
$\sum_{\sigma}$ is operated only on circulating particles with opposite
directions around $\bar{\psi}_{s}$. It should be noted that integration
$\oint\left(\cdots\right)d\vartheta_{c}$ is carried out surrounding
the rational surface $\bar{\psi}_{0}=\bar{\psi}_{s}$. The 0th order
vorticity equation is just 
\begin{equation}
\partial_{\vartheta_{0}}^{2}\delta\Psi^{\left(0\right)}=0,
\end{equation}
which gives the plain result that $\delta\Psi^{\left(0\right)}$ is
independent of $\vartheta_{0}$. From the Eq. (\ref{0th-quasi-neut}),
we can have a formal relation 
\begin{equation}
\delta\Phi^{\left(0\right)}=N_{0}\delta\Psi^{\left(0\right)},\label{0th-mhd}
\end{equation}
where $N_{0}=D_{1}/D_{2}$, 
\begin{align}
D_{1} & =\left(1+\frac{1}{\tau}\right)+\frac{T_{i}}{eN_{i}}\left\langle \frac{k_{\perp}}{k_{\vartheta}}\frac{\left(\frac{e}{m}QF_{0}\right)_{e}}{{\cal J}\omega_{be}}\frac{k_{\vartheta}/k_{\perp}}{\omega-\bar{\omega}_{de}}\left(\frac{\bar{\omega}_{de}}{\omega}-1\right)\right\rangle _{tr,\vartheta_{c}}\nonumber \\
 & +\frac{T_{i}}{eN_{i}}\left\langle \frac{k_{\perp}}{k_{\vartheta}}\frac{1}{{\cal J}\omega_{bi}}\frac{\partial\bar{\psi_{i}}}{\partial\psi}\left(\frac{e}{m}\right)_{i}\frac{QF_{0i}}{\omega}\frac{k_{\vartheta}}{k_{\perp}}\right\rangle _{circ,\vartheta_{c}}\nonumber \\
 & +\frac{T_{i}}{eN_{i}}\left\langle \frac{k_{\perp}}{k_{\vartheta}}\frac{\partial\bar{\psi_{i}}}{\partial\psi}\frac{e^{-i\tilde{Q}_{b}}}{2\pi{\cal J}\omega_{bi}}\sum_{l}e^{il\vartheta_{c}}\oint\frac{\left(\frac{e}{m}QF_{0}\right)_{i}\frac{k_{\vartheta}}{k_{\perp}}\left(\frac{\omega_{di}}{\omega}-1\right)e^{i\tilde{Q}_{b}\left(\vartheta_{c}^{\prime}\right)-il\vartheta_{c}^{\prime}}}{-\omega+\bar{\omega}_{di}+\left(l-n\bar{q}\right)\omega_{bi}}d\vartheta_{c}\right\rangle _{tr,\vartheta_{c}},
\end{align}
and
\begin{align*}
D_{2} & =\left(1+\frac{1}{\tau}\right)\frac{T_{i}}{eN_{i}}\left\langle \frac{k_{\perp}}{k_{\vartheta}}\frac{\left(\frac{e}{m}QF_{0}\right)_{e}}{{\cal J}\omega_{be}}\frac{k_{\vartheta}/k_{\perp}}{\omega-\bar{\omega}_{de}}\right\rangle _{tr,\vartheta_{c}}\\
 & +\frac{T_{i}}{eN_{i}}\left\langle \frac{k_{\perp}}{k_{\vartheta}}\frac{1}{{\cal J}\omega_{bi}}\frac{\partial\bar{\psi_{i}}}{\partial\psi}\left(\frac{e}{m}\right)_{i}\frac{QF_{0i}}{\omega}\frac{k_{\vartheta}}{k_{\perp}}\right\rangle _{circ,\vartheta_{c}}\\
 & +\frac{T_{i}}{eN_{i}}\left\langle \frac{k_{\perp}}{k_{\vartheta}}\frac{\partial\bar{\psi_{i}}}{\partial\psi}\frac{e^{-i\tilde{Q}_{b}}}{2\pi{\cal J}\omega_{bi}}\sum_{l}e^{il\vartheta_{c}}\oint\frac{\left(\frac{e}{m}QF_{0}\right)_{i}\frac{k_{\vartheta}}{k_{\perp}}e^{i\tilde{Q}_{b}\left(\vartheta_{c}^{\prime}\right)-il\vartheta_{c}^{\prime}}}{-\omega+\bar{\omega}_{di}+\left(l-n\bar{q}\right)\omega_{bi}}d\vartheta_{c}\right\rangle _{tr,\vartheta_{c}}
\end{align*}
As shown above, the ideal MHD condition, i.e. $\delta\Phi=\delta\Psi$,
is modified by trapped particles. The 1st order vorticity equation
is $\partial_{\vartheta_{0}}^{2}\delta\Psi^{\left(1\right)}=0$, which
gives the result that $\delta\Psi^{\left(1\right)}=0$. And $\delta\Phi^{\left(1\right)}=\sum_{l}\delta\Phi_{l}e^{il\vartheta_{0}}$
in general. The 1st kinetic equation for circulating ions can be given
as
\begin{equation}
\left(\omega_{b}\partial_{\vartheta_{c}}-i\omega\right)\delta K_{circ}^{\left(1\right)}=-\dot{\vartheta}\partial_{\vartheta_{1}}\delta K_{circ}^{\left(0\right)}+i\frac{k_{\vartheta}}{k_{\perp}}\left(\frac{e}{m}\right)QF_{0i}\left[\delta\Phi^{\left(1\right)}+\frac{\omega_{di}}{\omega}\delta\Phi^{\left(0\right)}\right].\label{1st-kic}
\end{equation}
The solution can be written as
\begin{align}
\delta K_{circ}^{\left(1\right)}= & -\sum_{l}\left(\frac{e}{m}\right)_{i}\frac{1}{\omega-l\omega_{bi}}\left(\frac{1}{2\pi}\sum_{l^{\prime}}\delta\Phi_{l^{\prime}}\oint\frac{k_{\vartheta}}{k_{\perp}}QF_{0i}e^{il^{\prime}\vartheta_{0}^{\prime}-il\vartheta_{c}^{\prime}}d\vartheta_{c}^{\prime}\right)e^{il\vartheta_{c}}\nonumber \\
 & -\sum_{l}\frac{1}{\omega\left(\omega-l\omega_{bi}\right)}\left(\frac{e}{m}\right)_{i}\delta\Phi^{\left(0\right)}\left(\frac{1}{2\pi}\oint\frac{k_{\vartheta}}{k_{\perp}}QF_{0i}\omega_{di}e^{-il\vartheta_{c}^{\prime}}d\vartheta_{c}^{\prime}\right)e^{il\vartheta_{c}}\nonumber \\
 & -\frac{i}{\omega}\partial_{\vartheta_{1}}\delta K_{circ}^{\left(0\right)}.\label{1st-kcir}
\end{align}
Then the 1st order quasi-neutrality of the l-th component can be cast
as
\begin{align}
\left(1+\frac{1}{\tau}\right)\delta\Phi_{l} & =\frac{T_{i}}{eN_{i}}\left\langle \frac{k_{\perp}}{k_{\vartheta}}\frac{e^{-il\vartheta_{0}}}{{\cal J}\omega_{bi}}\frac{\partial\bar{\psi}_{i}}{\partial\psi}\delta K_{circ}^{\left(1\right)}\right\rangle _{circ,\vartheta_{c}}\nonumber \\
 & +\frac{T_{i}}{eN_{i}}\left\langle \frac{k_{\perp}}{k_{\vartheta}}\frac{1-\delta_{l0}}{{\cal J}\omega_{bi}}e^{-il\vartheta_{0}}\frac{\partial\bar{\psi}_{i}}{\partial\psi}\delta K_{tr}^{\left(0\right)}\right\rangle _{tr,\vartheta_{c}},\label{1st-l-comp}
\end{align}
where $1-\delta_{l0}$ is introduced to avoid the double-counting
of the $l=0$ part of $\delta K_{tr}^{\left(0\right)}$ and $\delta_{l0}$
is the Kronecker symbol. If we cut off the harmonic to finite $l$,
the Eqs. (\ref{0th-mhd}), (\ref{1st-kcir}) and (\ref{1st-l-comp})
form a complete set of equations to give the relation
\begin{equation}
\delta\Phi_{l}=N_{l}\delta\Psi^{\left(0\right)},
\end{equation}
where $N_{l}=M_{ll^{\prime}}^{-1}X_{l^{\prime}}$, 
\begin{align*}
X_{l} & =-\frac{T_{i}N_{0}}{eN_{i}}\left\langle \frac{k_{\perp}}{k_{\vartheta}}\frac{e^{il\vartheta_{c}-il\vartheta_{0}}}{{\cal J}\omega_{bi}}\frac{\partial\bar{\psi}_{i}}{\partial\psi}\sum_{l}\frac{\left(\frac{e}{m}\right)_{i}\left(\frac{1}{2\pi}\oint\frac{k_{\vartheta}}{k_{\perp}}QF_{0i}\omega_{di}e^{-il\vartheta_{c}^{\prime}}d\vartheta_{c}^{\prime}\right)}{\omega\left(\omega-l\omega_{bi}\right)}\right\rangle _{circ,\vartheta_{c}}\\
 & -\frac{iT_{i}}{eN_{i}\omega\delta\Psi^{\left(0\right)}}\left\langle \frac{k_{\perp}}{k_{\vartheta}}\frac{e^{-il\vartheta_{0}}}{{\cal J}\omega_{bi}}\frac{\partial\bar{\psi}_{i}}{\partial\psi}\partial_{\vartheta_{1}}\delta K_{circ}^{\left(0\right)}\right\rangle _{circ,\vartheta_{c}}\delta_{l0}\\
 & \frac{T_{i}}{eN_{i}\delta\Psi^{\left(0\right)}}\left\langle \frac{k_{\perp}}{k_{\vartheta}}\frac{1-\delta_{l0}}{{\cal J}\omega_{bi}}e^{-il\vartheta_{0}}\frac{\partial\bar{\psi}_{i}}{\partial\psi}\delta K_{tr}^{\left(0\right)}\right\rangle _{tr,\vartheta_{c}},
\end{align*}
and
\begin{align}
M_{ll^{\prime}} & =\left(1+\frac{1}{\tau}\right)\delta_{ll^{\prime}}+\frac{T_{i}}{eN_{i}}\left\langle \frac{k_{\perp}}{k_{\vartheta}}\sum_{l^{\prime\prime}}\frac{e^{il^{\prime\prime}\vartheta_{c}-il\vartheta_{0}}}{{\cal J}\omega_{bi}}\frac{\partial\bar{\psi}_{i}}{\partial\psi}\left(\frac{e}{m}\right)_{i}\frac{\left(\frac{1}{2\pi}\oint\frac{k_{\vartheta}}{k_{\perp}}QF_{0i}e^{il^{\prime}\vartheta_{0}^{\prime}-il^{\prime\prime}\vartheta_{c}^{\prime}}d\vartheta_{c}^{\prime}\right)}{\omega-l^{\prime\prime}\omega_{bi}}\right\rangle _{circ,\vartheta_{c}}.
\end{align}
By averaging $\vartheta_{0}$, the 2nd order vorticity equation can
be obtained as
\begin{align}
 & \left(\frac{\partial^{2}}{\partial\vartheta_{1}^{2}}-\frac{1}{2\pi}\oint\frac{\partial_{\vartheta}^{2}\left|\nabla\psi\right|}{\left|\nabla\psi\right|}d\vartheta_{0}\right)\delta\Psi^{\left(0\right)}\nonumber \\
 & +\frac{4\pi}{c^{2}}\sum_{s}\left\langle \frac{e_{s}^{2}}{m_{s}}{\cal J}B^{2}\frac{v_{\perp}\omega}{k_{\perp}^{2}\omega_{bs}}\frac{\partial\bar{\psi}_{s}}{\partial\psi}J_{0}J_{1}QF_{0s}\right\rangle _{full,\vartheta_{c}}\delta\hat{B}_{\parallel}^{\left(0\right)}\nonumber \\
 & +\frac{4\pi}{c^{2}}\sum_{s}\frac{e_{s}^{2}}{m_{s}}\left\langle \frac{{\cal J}B^{2}}{k_{\perp}^{2}}J_{0}^{2}\frac{\omega_{ds}}{\omega_{bs}}\frac{\partial\bar{\psi}_{s}}{\partial\psi}QF_{0s}\right\rangle _{full,\vartheta_{c}}\delta\Psi^{\left(0\right)}\nonumber \\
 & -\frac{4\pi N_{0}}{c^{2}}\sum_{s}\frac{e_{s}^{2}}{m_{s}}\left\langle \frac{{\cal J}B^{2}}{k_{\perp}^{2}}\frac{\omega}{\omega_{bs}}\left(1-J_{0}^{2}\right)\frac{\partial\bar{\psi}_{s}}{\partial\psi}QF_{0s}\right\rangle \delta\Psi^{\left(0\right)}\nonumber \\
 & =\left\langle \frac{{\cal J}B^{2}}{\hat{\kappa}_{\perp}}\frac{4\pi e_{i}}{k_{\vartheta}^{2}c^{2}}\frac{\omega\omega_{di}}{\omega_{bi}}J_{0}\frac{\partial\bar{\psi}_{s}}{\partial\psi}\delta K_{circ}^{\left(1\right)}\right\rangle _{circ,\vartheta_{c}}\nonumber \\
 & +\sum_{s}\left\langle \frac{{\cal J}B^{2}}{\hat{\kappa}_{\perp}}\frac{4\pi e_{s}}{k_{\vartheta}^{2}c^{2}}\frac{\omega\omega_{ds}}{\omega_{bs}}J_{0}\frac{\partial\bar{\psi}_{s}}{\partial\psi}\delta K_{tr,s}^{\left(0\right)}\right\rangle _{tr,\vartheta_{c}}.\label{2nd-vor}
\end{align}
Here for the leading order terms, $J_{0}^{2}\simeq1$, $1-J_{0}^{2}\simeq k_{\perp}^{2}\rho_{Ls}^{2}/2$
and $J_{1}\simeq k_{\perp}\rho_{Ls}/2$. With Eqs. (\ref{0th-kcir})
, (\ref{0th-kte}), (\ref{0th-kti}), (\ref{0th-quasi-neut}), (\ref{1st-kcir})
and (\ref{1st-l-comp}), Eq. (\ref{2nd-vor}) can be solved, in general,
numerically provided that the information of particle orbit on the
grid of $\left({\cal E},\mu\right)$ and general magnetic geometry
data are given numerically. In this general model, we can consider
arbitrary magnetic configuration and ion orbit width\citep{white_hamiltonian_1984}. 

\section{An application to circular geometry}

\subsection{Reduced model}

In this application, we will only consider the leading order solution
without assuming well circulating and deeply trapped particles. For
the circular up-down symmetric configuration, the magnetic field $B=B_{0}\left(1-\epsilon\cos\theta\right)$,
where $\epsilon=r/R_{0}\sim\delta^{2}$ is the inverse aspect ratio.
Also, ${\cal J}B_{0}^{2}\simeq qR_{0}B_{0}$, $\tilde{Q}_{b}\simeq0$
and $\bar{\psi}_{s}\simeq\psi$ because of the small orbit width approximation.
Thus, for trapped ions, $\bar{q}\simeq0$. The guiding center motion
is governed by 
\begin{equation}
\dot{\theta_{s}}=\frac{v_{\parallel}}{qR_{0}},\label{theta_dot}
\end{equation}
\begin{equation}
\dot{\zeta_{s}}-q\dot{\theta_{s}}=\frac{q}{2rR_{0}\Omega_{s}}\left(v^{2}+v_{\parallel}^{2}\right)\cos\theta,
\end{equation}
\begin{equation}
q^{\prime}\dot{\psi_{s}}\vartheta_{1}=\frac{q}{2rR_{0}\Omega_{s}}\left(v^{2}+v_{\parallel}^{2}\right)s\vartheta_{1}\sin\theta,
\end{equation}
where $v_{\parallel}=\sigma\sqrt{2{\cal E}\left(1-\lambda B\right)}$,
$\lambda=\mu B_{0}/{\cal E}$ and $v=\sqrt{2{\cal E}}$. Then the
drift frequency can be rewritten by
\begin{equation}
\omega_{ds}=\frac{nq}{2rR_{0}\Omega_{s}}\left(v^{2}+v_{\parallel}^{2}\right)\left(\cos\theta+s\vartheta_{1}\sin\theta\right).
\end{equation}
Then we have the bouncing frequency for circulating particles with
$0<\lambda<1-\epsilon$ can be obtained as,
\begin{equation}
\omega_{bs}=\frac{\pi v\sqrt{2\epsilon\lambda}}{2qR_{0}}\frac{\kappa}{K\left(\kappa^{-1}\right)},
\end{equation}
where $K\left(\kappa\right)$ is the elliptic integral of the first
kind, $\kappa^{2}=\frac{1-\left(1-\epsilon\right)\lambda}{2\epsilon\lambda}$,
and $\kappa>1$ for circulating particles. For trapped particles with
$1-\epsilon<\lambda<1+\epsilon$, the bouncing frequency is
\begin{equation}
\omega_{bs}=\frac{\pi v\sqrt{2\epsilon\lambda}}{4qR_{0}}K^{-1}\left(\kappa\right),
\end{equation}
where $0<\kappa<1$ for trapped particles. The bouncing averaged drift
frequency for trapped particles is 
\begin{equation}
\bar{\omega}_{ds}=\frac{k_{\vartheta}{\cal E}}{R_{0}\Omega_{s}}\left(2-\lambda\right)\left[2\frac{E\left(\kappa\right)}{K\left(\kappa\right)}-1\right],
\end{equation}
where $E\left(\kappa\right)$ is the complete elliptic integrals of
the kinds. Also, the quantities related to wave numbers are given
as $k_{\vartheta}=nq/r$ and $\hat{\kappa}\simeq\sqrt{1+s^{2}\vartheta_{1}^{2}}$
in inertial layer.

\subsection{Canonical angle and pseudo-orthogonality}

Since the magnetic configuration is up-down symmetric, the canonical
angle can be defined piece-wisely. First, we consider the circulating
particles with $\sigma=1$. For $0\leq\vartheta_{0}<\pi$, 
\begin{equation}
\vartheta_{c}\left(\vartheta_{0},\sigma=1\right)=\vartheta_{circ}^{+}\left(\vartheta_{0}\right)=2\pi\frac{\int_{0}^{\vartheta_{0}}d\vartheta_{0}^{\prime}/\dot{\theta}}{\ointclockwise d\vartheta_{0}/\dot{\theta}}=\pi\frac{K\left(\theta/2,\kappa^{-1}\right)}{K\left(\kappa^{-1}\right)},
\end{equation}
where $K\left(x,\kappa\right)$ is the incomplete elliptic integral
for the first kind. And for $\pi\leq\vartheta_{0}<2\pi$, 
\begin{equation}
\vartheta_{c}\left(\vartheta_{0},\sigma=1\right)=2\pi-\vartheta_{c}^{+}\left(2\pi-\vartheta_{0}\right).
\end{equation}
For $\sigma=-1$, we have the reflection formula
\begin{equation}
\vartheta_{c}\left(\vartheta_{0},\sigma=-1\right)=2\pi-\vartheta_{c}\left(\vartheta_{0},\sigma=1\right),\label{can_angle-1}
\end{equation}
since the orbit width is neglected. For trapped particles, the canonical
angle is obtained piece wisely as
\begin{equation}
\theta_{c}\left(\vartheta_{0}\right)=\theta_{tr}^{+}\left(\vartheta_{0}\right)=2\pi\frac{\int_{0}^{\vartheta_{0}}d\vartheta_{0}^{\prime}/\dot{\theta}}{\ointclockwise d\vartheta_{0}/\dot{\theta}}=\frac{\pi K\left(x,\kappa\right)}{2K\left(\kappa\right)},\quad\textnormal{for }0\leq\theta_{0}\leq\theta_{b}\textnormal{ and }\dot{\theta}>0,
\end{equation}
where $sinx=\sin\left(\theta/2\right)/\sin\left(\theta_{b}/2\right)$.
With the up-down symmetry, the definition of canonical angle along
the rest parts of the orbit can be obtained as
\begin{equation}
\theta_{c}\left(\vartheta_{0}\right)=\begin{cases}
\pi-\theta_{tr}^{+}\left(\theta\right), & \textnormal{for }\theta_{b}>\theta_{0}\geq0\textnormal{ and }\dot{\theta}<0,\\
\pi+\theta_{tr}^{+}\left(2\pi-\theta\right), & \textnormal{for }2\pi>\theta_{0}\geq2\pi-\theta_{b}\textnormal{ and }\dot{\theta}<0,\\
2\pi-\theta_{tr}^{+}\left(2\pi-\theta\right), & \textnormal{for }2\pi-\theta_{b}<\theta_{0}<2\pi\textnormal{ and }\dot{\theta}>0.
\end{cases}
\end{equation}
With the definitions for $\vartheta_{c}$ and Eq. (\ref{theta_dot}),
the pseudo-orthogonality relation in bouncing average for both trapped
and circulating particles can be given as 
\begin{equation}
\oint\sin m\vartheta_{0}\cos l\vartheta_{c}d\vartheta_{c}=\oint\cos m\vartheta_{0}\sin l\vartheta_{c}d\vartheta_{c}=0,\label{pseudo-orth}
\end{equation}
where $m$ and $l$ are integers. Moreover, for trapped particles
and even $l$, we have 
\begin{equation}
\oint\sin m\vartheta_{0}\sin l\vartheta_{c}d\vartheta_{c}=\oint\cos m\vartheta_{0}\cos\left[\left(l+1\right)\vartheta_{c}\right]d\vartheta_{0}=0.\label{pseudo-orth-tr}
\end{equation}
As shown above, the Fourier components in $\vartheta_{0}$ and $\vartheta_{c}$,
by averaging over $\vartheta_{c}$, have the orthogonality similar
to the Fourier components only in $\vartheta_{0}$ . 

\subsection{Solutions for circular configuration}

As discussed above, Eq. (\ref{1st-l-comp}) needs to be solved for
finite $l$. And we assume that $\delta\Phi\simeq\delta\Phi^{\left(0\right)}+\delta\Phi_{s}\sin\vartheta_{0}$
and $\delta\Psi^{\left(1\right)}\simeq0$ \textit{a posteriori }for
the leading order solution. The 0th order solution of trapped ion
response in Eq. (\ref{0th-kti}) can be reduced to 
\begin{align}
\delta K_{tr,i}^{\left(0\right)} & =-\frac{1}{\omega-\bar{\omega}_{di}}\frac{e}{m}QF_{0i}\frac{k_{\vartheta}}{k_{\perp}}\left(\delta\Phi^{\left(0\right)}-\delta\Psi^{\left(0\right)}+\frac{\bar{\omega}_{di}}{\omega}\delta\Psi^{\left(0\right)}\right)\nonumber \\
 & -\frac{e}{m}QF_{0i}L\frac{\left(\omega-\bar{\omega}_{di}\right)k_{\vartheta}/k_{\perp}}{\left(\omega-\bar{\omega}_{di}\right)^{2}-\omega_{bi}^{2}}\left[\delta\Phi_{s}+\frac{k_{\vartheta}\Omega_{d}{\cal E}}{\omega{\cal E}_{t}}\left(2-\lambda\right)s\vartheta_{1}\delta\Psi^{\left(0\right)}\right]\sin\vartheta_{c}\nonumber \\
 & +\left[\cdots\right]\cos\vartheta_{c},
\end{align}
where $L\left(\epsilon,\lambda\right)=\frac{1}{\pi}\oint\sin\vartheta_{0}\sin\vartheta_{c}d\vartheta_{c}$
for trapped ions, $\Omega_{d}={\cal E}_{t}/\left(R_{0}\Omega_{i}\right)$,
and ${\cal E}_{t}=\frac{T_{i}}{m_{i}}$. Because of the pseudo-orthogonality
in Eq. (\ref{pseudo-orth}), the parts proportional to $\cos\vartheta_{c}$
are irrelevant to the leading order solution. And for trapped ions
the $l=\pm2$ can is approximately zero due to Eq. (\ref{pseudo-orth-tr}).
As we already assumed that the electron orbit width is negligible,
the solution of $\delta K_{tr,e}$ remains the same forms as in Eq.
(\ref{0th-kte}). Also, $\delta K_{cir}^{\left(0\right)}$ has the
same form in Eq. (\ref{0th-kcir}). The 0th order quasi-neutrality
equation can be written as
\begin{align}
\left(1+\frac{1}{\tau}\right)\left(\delta\Phi^{\left(0\right)}-\delta\Psi^{\left(0\right)}\right) & =-\frac{T_{i}}{m_{e}N_{i}}\left\langle \frac{QF_{0e}}{\omega-\bar{\omega}_{de}}\left[\frac{\bar{\omega}_{de}}{\omega}\delta\Psi^{\left(0\right)}+\delta\Phi^{\left(0\right)}-\delta\Psi^{\left(0\right)}\right]\right\rangle _{tr}\nonumber \\
-\frac{T_{i}}{m_{i}N_{i}}\left\langle \frac{QF_{0i}}{\omega}\left(\delta\Phi^{\left(0\right)}-\delta\Psi^{\left(0\right)}\right)\right\rangle _{circ} & -\frac{T_{i}}{m_{i}N_{i}}\left\langle \frac{QF_{0i}}{\omega-\bar{\omega}_{di}}\left(\delta\Phi^{\left(0\right)}-\delta\Psi^{\left(0\right)}+\frac{\bar{\omega}_{di}}{\omega}\delta\Psi^{\left(0\right)}\right)\right\rangle _{tr},\label{0th-qua-neu-cir}
\end{align}
where$N_{i}=\left\langle F_{0i}\right\rangle _{circ}+\left\langle F_{0i}\right\rangle _{tr}$,
$\left\langle \cdots\right\rangle _{tr}=\int_{0}^{\infty}\int_{1-\epsilon}^{1+\epsilon}\left(\cdots\right)\frac{\tau_{bs}{\cal E}}{qR_{0}}d\lambda d{\cal E}$
for trapped particles, $\left\langle \cdots\right\rangle _{circ}=2\int_{0}^{\infty}\int_{0}^{1-\epsilon}\left(\cdots\right)\frac{\tau_{bs}{\cal E}}{qR_{0}}d\lambda d{\cal E}$
for circulating particles, and $\tau_{bs}=\frac{2\pi}{\omega_{bs}}$.
By considering the $l=\pm1$ components, the 1st order solution of
$\delta K_{circ}^{\left(1\right)}$ in Eq. (\ref{1st-kcir}) can be
reduced to 
\begin{align}
\delta K_{circ}^{\left(1\right)} & =-\left(\frac{e}{m}\right)\frac{1}{\omega^{2}-\omega_{bi}^{2}}\frac{k_{\vartheta}}{k_{\perp}}QF_{0i}M\left[\omega\delta\Phi_{s}+\frac{{\cal E}}{{\cal E}_{t}}k_{\vartheta}\Omega_{d}\left(2-\lambda\right)s\vartheta_{1}\delta\Psi^{\left(0\right)}\right]\sin\vartheta_{c}\nonumber \\
 & +\left[\cdots\right]\cos\vartheta_{c},
\end{align}
where $M\left(\epsilon,\lambda,\sigma\right)=\frac{1}{\pi}\oint\sin\vartheta_{0}\sin\vartheta_{c}d\vartheta_{c}$
for circulating ions. In later subsection, we will show that the approximation
$\vartheta_{c}\simeq\vartheta_{0}$is valid for the majority of circulating
ions. It should be noted that $\delta K_{circ}^{\left(1\right)}$
is symmetric with respect to $\sigma=\pm1$ according to Eq. (\ref{can_angle-1}).
And the 1st order quasi-neutrality equation of the $\delta\Phi_{s}$
part is 
\begin{align}
\left(1+\frac{1}{\tau}\right)\delta\Phi_{s} & =-\frac{T_{i}}{m_{i}N_{i}}\left\langle \frac{\left(\omega-\bar{\omega}_{d}\right)L^{2}QF_{0i}}{\left(\omega-\bar{\omega}_{di}\right)^{2}-\omega_{bi}^{2}}\left[\delta\Phi_{s}+\frac{k_{\vartheta}\Omega_{d}}{\omega}\frac{{\cal E}}{{\cal E}_{t}}\left(2-\lambda\right)s\vartheta_{1}\delta\Psi^{\left(0\right)}\right]\right\rangle _{tr}\nonumber \\
 & -\frac{T_{i}}{m_{i}N_{i}}\left\langle \frac{M^{2}QF_{0i}}{\omega^{2}-\omega_{bi}^{2}}\left[\omega\delta\Phi_{s}+k_{\vartheta}\Omega_{d}\frac{{\cal E}}{{\cal E}_{t}}\left(2-\lambda\right)s\vartheta_{1}\delta\Psi^{\left(0\right)}\right]\right\rangle _{circ}.
\end{align}
 Then we directly have $N_{0}=D_{1}/D_{2}$ and 
\begin{equation}
\delta\Phi_{s}=\frac{k_{\vartheta}s\vartheta_{1}\Omega_{d}}{\omega}N_{s}\delta\Psi^{\left(0\right)},\label{sine-cir}
\end{equation}
where $N_{s}=D_{3}/D_{4}$,
\begin{align}
D_{1} & =\left(1+\frac{1}{\tau}\right)-\frac{T_{i}}{m_{e}N_{i}}\left\langle \frac{QF_{0e}}{\omega-\bar{\omega}_{de}}\left(\frac{\bar{\omega}_{de}}{\omega}-1\right)\right\rangle _{tr}\nonumber \\
 & +\frac{T_{i}}{m_{i}N_{i}}\left\langle \frac{QF_{0i}}{\omega}\right\rangle _{circ}-\frac{T_{i}}{m_{i}N_{i}}\left\langle \frac{QF_{0i}}{\omega-\bar{\omega}_{di}}\left(\frac{\bar{\omega}_{di}}{\omega}-1\right)\right\rangle _{tr},
\end{align}
\begin{align}
D_{2} & =\left(1+\frac{1}{\tau}\right)+\frac{T_{i}}{m_{e}N_{i}}\left\langle \frac{QF_{0e}}{\omega-\bar{\omega}_{de}}\right\rangle _{tr}\nonumber \\
 & +\frac{T_{i}}{m_{i}N_{i}}\left\langle \frac{QF_{0i}}{\omega}\right\rangle _{circ}+\frac{T_{i}}{m_{i}N_{i}}\left\langle \frac{QF_{0i}}{\omega-\bar{\omega}_{di}}\right\rangle _{tr},
\end{align}
\begin{align}
D_{3} & =-\frac{T_{i}}{m_{i}N_{i}}\left\langle \frac{\left(\omega-\bar{\omega}_{di}\right)L^{2}QF_{0i}}{\left(\omega-\bar{\omega}_{di}\right)^{2}-\omega_{bi}^{2}}\frac{{\cal E}}{{\cal E}_{t}}\left(2-\lambda\right)\right\rangle _{tr}\nonumber \\
 & -\frac{T_{i}}{m_{i}N_{i}}\left\langle \frac{M^{2}\omega QF_{0i}}{\omega^{2}-\omega_{bi}^{2}}\frac{{\cal E}}{{\cal E}_{t}}\left(2-\lambda\right)\right\rangle _{circ},
\end{align}
and 
\begin{align}
D_{4} & =\left(1+\frac{1}{\tau}\right)+\frac{T_{i}}{m_{i}N_{i}}\left\langle \frac{M^{2}\omega QF_{0i}}{\omega^{2}-\omega_{bi}^{2}}\right\rangle _{circ}\nonumber \\
 & +\frac{T_{i}}{m_{i}N_{i}}\left\langle \frac{\left(\omega-\bar{\omega}_{di}\right)L^{2}QF_{0i}}{\left(\omega-\bar{\omega}_{di}\right)^{2}-\omega_{bi}^{2}}\right\rangle _{tr}.
\end{align}
Finally, the second order vorticity equation is obtained as
\begin{equation}
\frac{\partial^{2}\delta\Psi^{\left(0\right)}}{\partial\vartheta_{1}^{2}}+\Lambda^{2}\delta\Psi^{\left(0\right)}=0,
\end{equation}
where $\Lambda^{2}=\Lambda_{fl}^{2}+\Lambda_{circ}^{2}+\Lambda_{tr}^{2}$,
$I\left({\cal E},\lambda\right)=\frac{{\cal E}}{{\cal E}_{t}}\left(2-\lambda\right)$,
$\left\langle \cdots\right\rangle _{full}=\left\langle \cdots\right\rangle _{circ}+\left\langle \cdots\right\rangle _{tr}$,
\begin{equation}
\Lambda_{fl}^{2}=-\frac{8\pi^{2}\omega}{B_{0}^{2}}q^{2}R_{0}^{2}N_{0}\sum_{s}\left\langle m_{s}QF_{0s}{\cal E}\lambda^{2}\right\rangle _{full},\label{lambda_fl}
\end{equation}
 
\begin{equation}
\Lambda_{circ}^{2}=\frac{2\pi\omega e^{2}}{m_{i}c^{2}}q^{2}R_{0}^{2}\left\langle \frac{QF_{0i}M^{2}\Omega_{d}^{2}}{\omega^{2}-\omega_{bi}^{2}}\left(N_{s}I+I^{2}\right)\right\rangle _{circ},\label{lambda-circ}
\end{equation}
and 
\begin{equation}
\Lambda_{tr}^{2}=\frac{2\pi\omega e^{2}}{m_{i}c^{2}}q^{2}R_{0}^{2}\left\langle \frac{\left(\omega-\bar{\omega}_{di}\right)QF_{0i}L^{2}}{\left(\omega-\bar{\omega}_{di}\right)^{2}-\omega_{bi}^{2}}\frac{\Omega_{d}^{2}}{\omega}\left(N_{s}I+I^{2}\right)\right\rangle _{tr}.\label{lambda-tr}
\end{equation}
The MPC and MFC terms is of order $O\left(\delta^{3}\right)$ due
to the pseudo-orthogonal relation in Eq. (\ref{pseudo-orth}). Now
the inertia term $\Lambda^{2}$ contains the corrections from the
contribution of particles near circulating/trapped separatrix. As
shown in the results above, the particles near the barely circulating/trapped
sepparatrix can modify the inertia term. The detailed expressions
with Maxwellian equilibrium distribution are given in the appendix.

\subsection{Limiting case}

\begin{figure}
\begin{centering}
\includegraphics[scale=0.5]{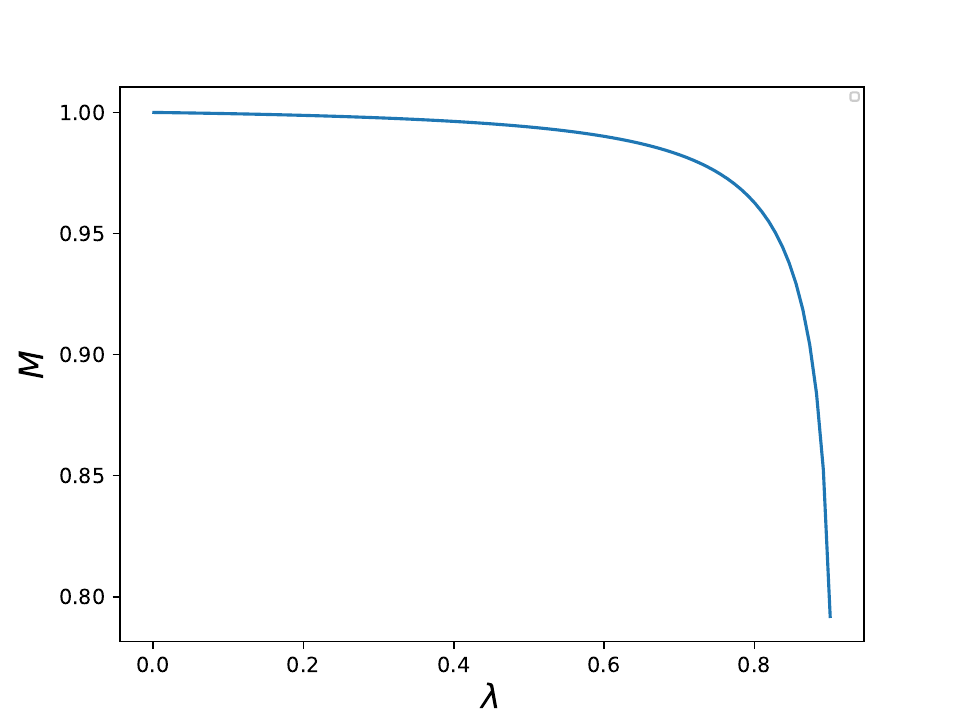}
\par\end{centering}
\caption{$M$ function for $\sigma=1$ and reverse aspect ratio $\epsilon=0.1$}
\label{M-function}
\end{figure}

\begin{figure}
\begin{centering}
\includegraphics[scale=0.5]{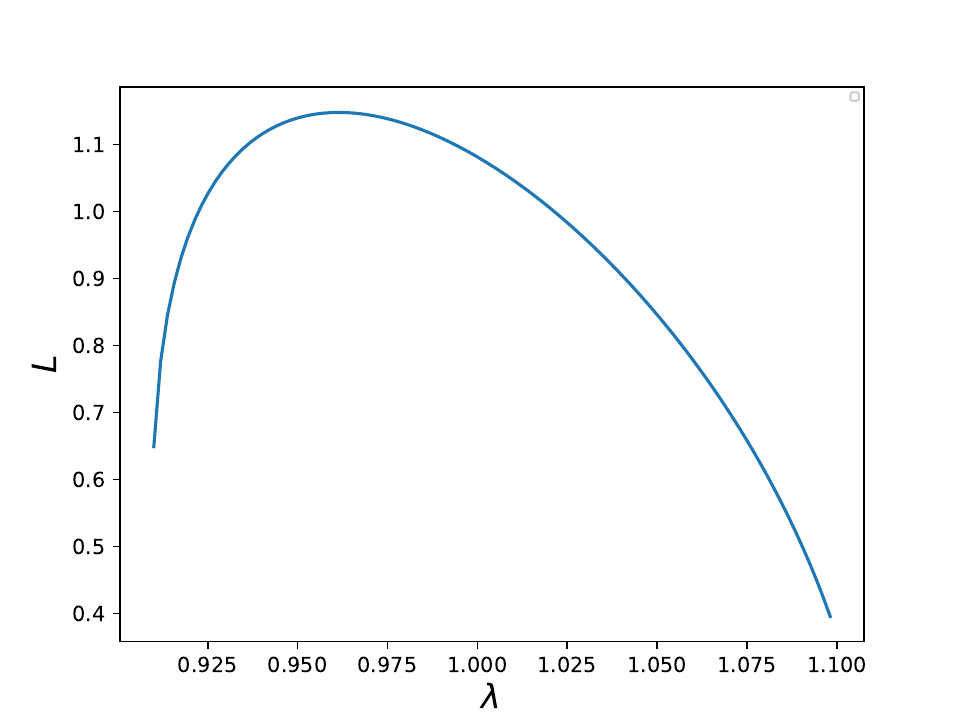}
\includegraphics[scale=0.5]{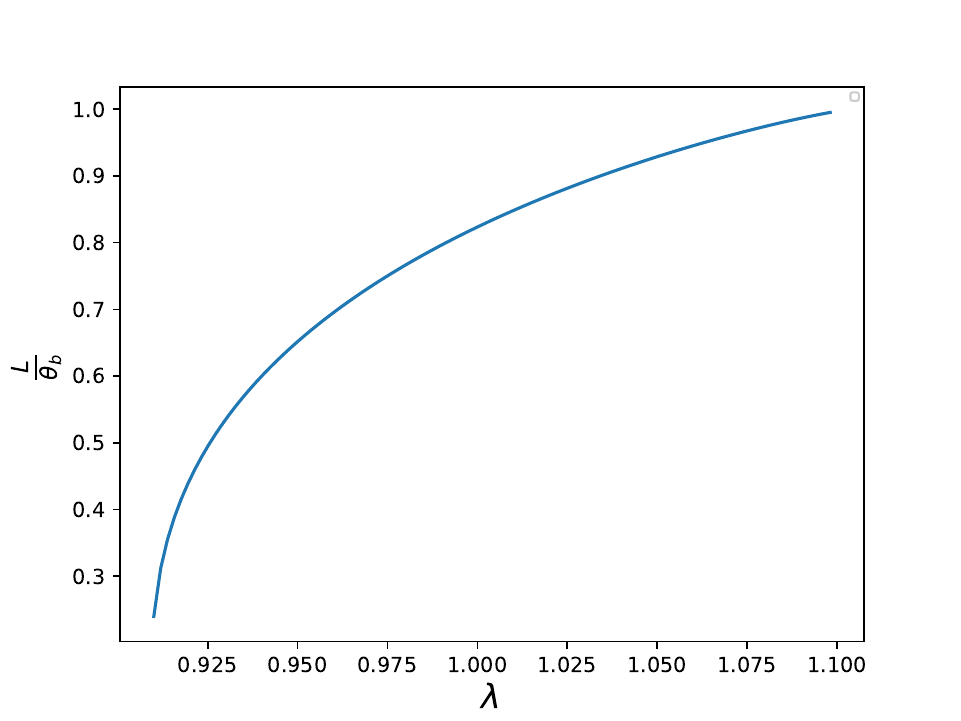}
\par\end{centering}
\caption{$L$ and $L/\theta_{b}$ functions for reverse aspect ratio $\epsilon=0.1$}

\label{L-function-theta-0}
\end{figure}

In this subsection, we will analyze the weight functions, i.e. $M$
and $L$ to demonstrate how the results are related to those in previous
researches\citep{chavdarovski_effects_2009}. If the circulating particles
are approximately taken to be well circulating, i.e. $\lambda\ll1$,
the canonical angle is then 
\begin{equation}
\vartheta_{c}\left(\vartheta_{0},\sigma=\pm1\right)=2\pi\delta_{-1,\sigma}+\sigma\vartheta_{0}.
\end{equation}
And the parallel velocity $v_{\parallel}$ can be treated approximately
independent of $\vartheta_{0}$. Then the weight term $M$ can be
given as 
\begin{equation}
M\left(\sigma=1\right)\simeq\frac{1}{\pi}\int_{0}^{2\pi}\sin\vartheta_{0}\sin\vartheta_{0}d\vartheta_{0}=1.\label{limit-weight-circ}
\end{equation}
And the velocity space integral is 
\begin{equation}
\left\langle \cdots\right\rangle _{circ}\simeq4\pi\int_{0}^{\infty}\int_{0}^{1-\epsilon}\left(\cdots\right)\sqrt{\frac{{\cal E}}{2\left(1-\lambda\right)}}d\lambda d{\cal E},\label{v-int-circ}
\end{equation}
For the trapped particles, by assuming trapped particles are deeply
trapped, we have $\theta_{b}\simeq2\kappa$, $\vartheta_{0}\simeq\theta_{b}\sin\vartheta_{c}$,
$v_{\parallel}\simeq\theta_{b}\omega_{b}qR_{0}\cos\vartheta_{c}$
and $\omega_{b}=\sqrt{\epsilon\mu B_{0}}/qR_{0}$. Then the weight
term $L$ and the velocity space integral $\left\langle \cdots\right\rangle _{tr}$
are given as 
\begin{equation}
L=\frac{4}{\pi}\int_{0}^{\pi/2}\sin\vartheta_{c}\sin\vartheta_{0}d\vartheta_{c}\simeq\theta_{b},\label{limit-weight-tr}
\end{equation}
 and 
\begin{equation}
\left\langle \cdots\right\rangle _{tr}\simeq\sqrt{\epsilon}4\pi\int_{0}^{\infty}\left[\cdots\right]{\cal E}^{1/2}d{\cal E}.\label{v-int-tr}
\end{equation}
Here the approximation $\sin\vartheta_{0}\simeq\theta_{b}\sin\vartheta_{c}$
has been used. As shown in Fig. \ref{M-function}, $M\simeq1$ for
except for the circulating ions near circulating/trapped separatrix.
And $\vartheta_{c}\simeq\vartheta_{0}$ is a good approximation for
the majority of circulating ions. In Figure. \ref{L-function-theta-0},
it is illustrated that the value of $L/\theta_{b}$ approaches $1$
for deeply trapped particles. Moreover, the value of the $L$ function
peaks at the medium trapped region, which indicates that the contribution
from the medium trapped particles is bigger than that of the deeply
trapped particles. Finally, with Eqs. (\ref{limit-weight-circ}),
(\ref{v-int-circ}), (\ref{limit-weight-tr}) and (\ref{v-int-tr}),
the results in Subsection 4.3 directly go back the previous results
with well-circulating and deeply trapped particles\citep{chavdarovski_effects_2009,chavdarovski_analytic_2014}.

\section{Summary and discussion}

The results in this work can be summarized as follows:
\begin{enumerate}
\item We develop a comprehensive kinetic model with full orbit effects,
including the motion of guiding centers in general geometric configurations. 
\item It is demonstrated that the vorticity equation in the inertial layer
can be obtained immediately with ion orbit data. 
\item we present a simplified model that incorporates full neoclassical
effects in the context of circular magnetic geometry, under the assumption
of small ion orbit width.
\item In the scenario where ions are assumed to be effectively circulating
or deeply trapped, the outcomes align with those of previous research
studies\citep{chavdarovski_effects_2009}.
\end{enumerate}
With the models developed in this work, more precise numerical calculations
can be carried out and serve as an intermediate model between gyrokinetic
simulations and experimental observations. By applying the general
model, we can study the geometric effect such as triangularity and
elongation etc. and the finite orbit width effect from circulating
and trapped ions. A quick calculation can be conducted with the simplified
model by neglecting ion orbit width and treating the magnetic geometry
as circular configuration approximately. Moreover, as demonstrated
in the preceding results, particles located in the vicinity of the
barely circulating/trapped separatrix can be included in the present
model. Thus the theoretical model in this work fills in the blank
areas of previous studies. Results from both models we developed can
be used as benching mark for simulations as well as prediction for
experiments. Moreover, the models presented in this paper can also
be applied to other problems involving two scale analysis in the inertial
layer such as GAM. The future plan based on the model in this work
will be conducting numerical calculations for both general and circular
geometry\citep{falessi_2023nonlinear}.

This work was supported by the Italian Ministry for Foreign Affairs
and International Cooperation Project under Grant No. CN23GR02. 

\appendix

\section*{Appendix A}

\setcounter{equation}{0}
\renewcommand\theequation{A\arabic{equation}}

In this appendix, the detailed expressions for Maxwellian distribution
function are given. The Maxwellian distribution is 
\begin{equation}
F_{0s}=\frac{n_{s}}{\left(2\pi{\cal E}_{ts}\right)^{3/2}}e^{-\frac{{\cal E}}{{\cal E}_{ts}}},
\end{equation}
where ${\cal E}_{ts}=T_{s}/m_{s}$. Then we have 
\begin{align}
D_{1} & =\left(1+\frac{1}{\tau}\right)-\frac{n_{i}}{\sqrt{2}\pi N_{i}}\int_{0}^{1-\epsilon}\frac{K\left(\kappa^{-1}\right)}{\kappa\sqrt{\epsilon\lambda}}\left(1-\frac{\omega_{*ni}}{\omega}\right)d\lambda\nonumber \\
 & -\frac{\sqrt{2}n_{e}}{\pi\tau N_{i}}\left(1-\frac{\omega_{*ne}}{\omega}+\frac{3}{2}\frac{\omega_{*Te}}{\omega}\right)\int_{1-\epsilon}^{1+\epsilon}\frac{K\left(\kappa\right)}{\sqrt{\epsilon\lambda}}\left[Z_{4}\left(\sqrt{\frac{\omega}{\bar{\omega}_{De}}}\right)-\frac{\omega}{\bar{\omega}_{De}}Z_{2}\left(\sqrt{\frac{\omega}{\bar{\omega}_{De}}}\right)\right]\sqrt{\frac{\bar{\omega}_{De}}{\omega}}d\lambda\nonumber \\
 & +\frac{\sqrt{2}n_{e}}{\pi\tau N_{i}}\frac{\omega_{*Te}}{\omega}\int_{1-\epsilon}^{1+\epsilon}\frac{K\left(\kappa\right)}{\sqrt{\epsilon\lambda}}\left[Z_{6}\left(\sqrt{\frac{\omega}{\bar{\omega}_{De}}}\right)-\frac{\omega}{\bar{\omega}_{De}}Z_{4}\left(\sqrt{\frac{\omega}{\bar{\omega}_{De}}}\right)\right]\sqrt{\frac{\bar{\omega}_{De}}{\omega}}d\lambda\nonumber \\
 & -\frac{\sqrt{2}n_{i}}{\pi N_{i}}\left(1-\frac{\omega_{*ni}}{\omega}+\frac{3}{2}\frac{\omega_{*Ti}}{\omega}\right)\int_{1-\epsilon}^{1+\epsilon}\frac{K\left(\kappa\right)}{\sqrt{\epsilon\lambda}}\sqrt{\frac{\bar{\omega}_{Di}}{\omega}}\left[Z_{4}\left(\sqrt{\frac{\omega}{\bar{\omega}_{Di}}}\right)-\frac{\omega}{\bar{\omega}_{Di}}Z_{2}\left(\sqrt{\frac{\omega}{\bar{\omega}_{Di}}}\right)\right]d\lambda\nonumber \\
 & +\frac{\sqrt{2}n_{i}}{\pi N_{i}}\frac{\omega_{*Ti}}{\omega}\int_{1-\epsilon}^{1+\epsilon}\frac{K\left(\kappa\right)}{\sqrt{\epsilon\lambda}}\sqrt{\frac{\bar{\omega}_{Di}}{\omega}}\left[Z_{6}\left(\sqrt{\frac{\omega}{\bar{\omega}_{Di}}}\right)-\frac{\omega}{\bar{\omega}_{De}}Z_{4}\left(\sqrt{\frac{\omega}{\bar{\omega}_{Di}}}\right)\right]d\lambda,
\end{align}
\begin{align}
D_{2} & =\left(1+\frac{1}{\tau}\right)+\frac{\sqrt{2}n_{e}}{\pi\tau N_{i}}\left(1-\frac{\omega_{*ne}}{\omega}+\frac{3}{2}\frac{\omega_{*Te}}{\omega}\right)\int_{1-\epsilon}^{1+\epsilon}\frac{K\left(\kappa\right)}{\sqrt{\epsilon\lambda}}\sqrt{\frac{\omega}{\bar{\omega}_{De}}}Z_{2}\left(\sqrt{\frac{\omega}{\bar{\omega}_{De}}}\right)d\lambda\nonumber \\
 & -\frac{\sqrt{2}n_{e}}{\pi\tau N_{i}}\frac{\omega_{*Te}}{\omega}\int_{1-\epsilon}^{1+\epsilon}\frac{K\left(\kappa\right)}{\sqrt{\epsilon\lambda}}\sqrt{\frac{\omega}{\bar{\omega}_{De}}}Z_{4}\left(\sqrt{\frac{\omega}{\bar{\omega}_{De}}}\right)d\lambda\nonumber \\
 & +\frac{\sqrt{2}n_{i}}{\pi N_{i}}\left(1-\frac{\omega_{*ni}}{\omega}+\frac{3\omega_{*Ti}}{2\omega}\right)\int_{1-\epsilon}^{1+\epsilon}\frac{K\left(\kappa\right)}{\sqrt{\epsilon\lambda}}\sqrt{\frac{\omega}{\bar{\omega}_{Di}}}Z_{2}\left(\sqrt{\frac{\omega}{\bar{\omega}_{Di}}}\right)d\lambda\nonumber \\
 & -\frac{\sqrt{2}n_{i}}{\pi N_{i}}\frac{\omega_{*Ti}}{\omega}\int_{1-\epsilon}^{1+\epsilon}\frac{K\left(\kappa\right)}{\sqrt{\epsilon\lambda}}\sqrt{\frac{\omega}{\bar{\omega}_{Di}}}Z_{4}\left(\sqrt{\frac{\omega}{\bar{\omega}_{Di}}}\right)d\lambda\nonumber \\
 & -\frac{n_{i}}{\sqrt{2}\pi N_{i}}\int_{0}^{1-\epsilon}\frac{K\left(\kappa^{-1}\right)}{\kappa\sqrt{\epsilon\lambda}}\left(1-\frac{\omega_{*ni}}{\omega}\right)d\lambda,
\end{align}
\begin{equation}
N_{i}=\frac{n_{i}}{\sqrt{2}\pi}\left(\int_{0}^{1-\epsilon}\frac{K\left(\kappa^{-1}\right)}{\kappa\sqrt{\epsilon\lambda}}d\lambda+\int_{1-\epsilon}^{1+\epsilon}\frac{K\left(\kappa\right)}{\sqrt{\epsilon\lambda}}d\lambda\right),
\end{equation}
where 
\begin{equation}
Z_{n}\left(x\right)=\frac{2x}{\sqrt{\pi}}\int_{0}^{\infty}\frac{t^{n}e^{-t^{2}}}{t^{2}-x^{2}}dt.
\end{equation}
For even $n$, $Z_{n}\left(x\right)=xZ_{n-1}\left(x\right)$. For
odd $n$, $Z_{n}\left(x\right)=xZ_{n-1}\left(x\right)+\Gamma\left(n/2\right)/\sqrt{\pi}$.
$Z_{0}\left(x\right)$ is the so-called plasma dispersion function.
And $\Gamma\left(x\right)$ is the Gamma function. Here the functions
in $\kappa$ are treated as functions in $\lambda$ for simplicity
of notation. Then we have 
\begin{align}
D_{3} & =\frac{\sqrt{2}n_{i}}{\pi N_{i}}\left(1-\frac{\omega_{*ni}}{\omega}+\frac{3}{2}\frac{\omega_{*Ti}}{\omega}\right)\int_{1-\epsilon}^{1+\epsilon}\frac{\omega}{\bar{\omega}_{Di}}\frac{L^{2}K\left(\kappa\right)}{\sqrt{\epsilon\lambda}}\left(2-\lambda\right)\left(\frac{\omega}{\bar{\omega}_{Di}}G_{4}-G_{6}\right)d\lambda\nonumber \\
 & -\frac{\sqrt{2}n_{i}}{\pi N_{i}}\frac{\omega_{*Ti}}{\omega}\int_{1-\epsilon}^{1+\epsilon}\frac{\omega}{\bar{\omega}_{Di}}\frac{L^{2}K\left(\kappa\right)}{\sqrt{\epsilon\lambda}}\left(2-\lambda\right)\left(\frac{\omega}{\bar{\omega}_{Di}}G_{6}-G_{8}\right)d\lambda\nonumber \\
 & -\frac{\sqrt{2}n_{i}}{\pi N_{i}}\left(1-\frac{\omega_{*ni}}{\omega}+\frac{3}{2}\frac{\omega_{*Ti}}{\omega}\right)\int_{0}^{1-\epsilon}\frac{M^{2}K\left(\kappa^{-1}\right)}{\kappa\sqrt{\epsilon\lambda}}\left(2-\lambda\right)\frac{\omega}{\omega_{Bi}}Z_{4}\left(\frac{\omega}{\omega_{Bi}}\right)d\lambda\nonumber \\
 & +\frac{\sqrt{2}n_{i}}{\pi N_{i}}\frac{\omega_{*Ti}}{\omega}\int_{0}^{1-\epsilon}\frac{M^{2}K\left(\kappa^{-1}\right)}{\kappa\sqrt{\epsilon\lambda}}\left(2-\lambda\right)\frac{\omega}{\omega_{Bi}}Z_{6}\left(\frac{\omega}{\omega_{Bi}}\right)d\lambda,
\end{align}
\begin{align}
D_{4} & =\left(1+\frac{1}{\tau}\right)-\frac{\sqrt{2}n_{i}}{\pi N_{i}}\left(1-\frac{\omega_{*ni}}{\omega}+\frac{3}{2}\frac{\omega_{*Ti}}{\omega}\right)\int_{1-\epsilon}^{1+\epsilon}\frac{\omega}{\bar{\omega}_{Di}}\frac{L^{2}K\left(\kappa\right)}{\sqrt{\epsilon\lambda}}\left(\frac{\omega}{\bar{\omega}_{Di}}G_{2}-G_{4}\right)d\lambda\nonumber \\
 & +\frac{\sqrt{2}n_{i}}{\pi N_{i}}\frac{\omega_{*Ti}}{\omega}\int_{1-\epsilon}^{1+\epsilon}\frac{\omega}{\bar{\omega}_{Di}}\frac{L^{2}K\left(\kappa\right)}{\sqrt{\epsilon\lambda}}\left(\frac{\omega}{\bar{\omega}_{Di}}G_{4}-G_{6}\right)d\lambda\nonumber \\
 & +\frac{\sqrt{2}n_{i}}{\pi N_{i}}\left(1-\frac{\omega_{*ni}}{\omega}+\frac{3}{2}\frac{\omega_{*Ti}}{\omega}\right)\int_{0}^{1-\epsilon}\frac{M^{2}K\left(\kappa^{-1}\right)}{\kappa\sqrt{\epsilon\lambda}}\frac{\omega}{\omega_{Bi}}Z_{2}\left(\frac{\omega}{\omega_{Bi}}\right)d\lambda\nonumber \\
 & -\frac{\sqrt{2}n_{i}}{\pi N_{i}}\frac{\omega_{*Ti}}{\omega}\int_{0}^{1-\epsilon}\frac{M^{2}K\left(\kappa^{-1}\right)}{\kappa\sqrt{\epsilon\lambda}}\left(2-\lambda\right)\frac{\omega}{\omega_{Bi}}Z_{4}\left(\frac{\omega}{\omega_{Bi}}\right)d\lambda,
\end{align}
where 
\begin{align}
G_{n} & =\frac{1}{\sqrt{\pi}}\int_{0}^{\infty}\frac{2t^{n}e^{-t^{2}}}{\left(\omega/\bar{\omega}_{Di}-t^{2}\right)^{2}-\omega_{Bi}^{2}t^{2}/\bar{\omega}_{Di}^{2}}dt\nonumber \\
 & =\frac{1}{\Omega_{1}^{2}-\Omega_{2}^{2}}\left[\frac{1}{\Omega_{1}}Z_{n}\left(\Omega_{1}\right)-\frac{1}{\Omega_{2}}Z_{n}\left(\Omega_{2}\right)\right],
\end{align}
\begin{equation}
\Omega_{1}=\frac{\frac{\omega_{Bi}}{\bar{\omega}_{Di}}+\sqrt{\frac{4\omega}{\bar{\omega}_{Di}}+\frac{\omega_{Bi}^{2}}{\bar{\omega}_{Di}^{2}}}}{2},
\end{equation}
and
\begin{equation}
\Omega_{2}=\frac{-\frac{\omega_{Bi}}{\bar{\omega}_{Di}}+\sqrt{\frac{4\omega}{\bar{\omega}_{Di}}+\frac{\omega_{Bi}^{2}}{\bar{\omega}_{Di}^{2}}}}{2}.
\end{equation}
Thus, $N_{0}=D_{1}/D_{2}$ and $N_{s}=D_{3}/D_{4}$ are given. For
the inertia terms, we have
\begin{equation}
\Lambda_{fl}^{2}=\frac{\omega\left(\omega-\omega_{*Pi}\right)}{2\pi\omega_{A}^{2}}N_{0}\left(\int_{0}^{1-\epsilon}\lambda^{1/2}\frac{3\sqrt{2}K\left(\kappa^{-1}\right)}{2\kappa\sqrt{\epsilon}}d\lambda+\int_{1-\epsilon}^{1+\epsilon}\frac{3\sqrt{2}\lambda^{1/2}K\left(\kappa\right)}{2\sqrt{\epsilon}}d\lambda\right),
\end{equation}
\begin{align}
\Lambda_{circ}^{2} & =\frac{\sqrt{2}\omega_{ti}^{2}}{4\pi\omega_{A}^{2}}q^{2}\omega\left(1-\frac{\omega_{*ni}}{\omega}+\frac{3}{2}\frac{\omega_{*Ti}}{\omega}\right)\int_{0}^{1-\epsilon}\frac{K\left(\kappa^{-1}\right)M^{2}}{\omega_{Bi}\kappa\sqrt{\epsilon\lambda}}\left(2-\lambda\right)\left[N_{s}Z_{4}\left(\frac{\omega}{\omega_{Bi}}\right)+\left(2-\lambda\right)Z_{6}\left(\frac{\omega}{\omega_{Bi}}\right)\right]d\lambda\nonumber \\
 & -\frac{\sqrt{2}\omega_{ti}^{2}}{4\pi\omega_{A}^{2}}q^{2}\omega_{*Ti}\int_{0}^{1-\epsilon}\frac{K\left(\kappa^{-1}\right)M^{2}}{\omega_{Bi}\kappa\sqrt{\epsilon\lambda}}\left(2-\lambda\right)\left[N_{s}Z_{6}\left(\frac{\omega}{\omega_{Bi}}\right)+\left(2-\lambda\right)Z_{8}\left(\frac{\omega}{\omega_{Bi}}\right)\right]d\lambda,
\end{align}
and
\begin{align}
\Lambda_{tr}^{2} & =-\frac{\sqrt{2}\omega_{ti}^{2}}{4\pi\omega_{A}^{2}}q^{2}\left(1-\frac{\omega_{*ni}}{\omega}+\frac{3}{2}\frac{\omega_{*Ti}}{\omega}\right)\int_{1-\epsilon}^{1+\epsilon}\frac{\omega^{2}}{\bar{\omega}_{Di}^{2}}\frac{K\left(\kappa\right)L^{2}}{\sqrt{\epsilon\lambda}}\left(2-\lambda\right)\left[N_{s}G_{4}+G_{6}\left(2-\lambda\right)\right]d\lambda\nonumber \\
 & +\frac{\sqrt{2}\omega_{ti}^{2}}{4\pi\omega_{A}^{2}}q^{2}\frac{\omega_{*Ti}}{\omega}\int_{1-\epsilon}^{1+\epsilon}\frac{K\left(\kappa\right)\left(2-\lambda\right)L^{2}}{\sqrt{\epsilon\lambda}}\frac{\omega^{2}}{\bar{\omega}_{Di}^{2}}\left[\left(2-\lambda\right)G_{8}+N_{s}G_{6}\right]d\lambda\nonumber \\
 & +\frac{\sqrt{2}\omega_{ti}^{2}}{4\pi\omega_{A}^{2}}q^{2}\left(1-\frac{\omega_{*ni}}{\omega}+\frac{3}{2}\frac{\omega_{*Ti}}{\omega}\right)\int_{1-\epsilon}^{1+\epsilon}\frac{K\left(\kappa\right)\left(2-\lambda\right)L^{2}}{\sqrt{\epsilon\lambda}}\frac{\omega}{\bar{\omega}_{Di}}\left[\left(2-\lambda\right)G_{8}+N_{s}G_{6}\right]d\lambda\nonumber \\
 & -\frac{\sqrt{2}\omega_{ti}^{2}}{4\pi\omega_{A}^{2}}q^{2}\frac{\omega_{*Ti}}{\omega}\int_{1-\epsilon}^{1+\epsilon}\frac{K\left(\kappa\right)\left(2-\lambda\right)L^{2}}{\sqrt{\epsilon\lambda}}\frac{\omega}{\bar{\omega}_{Di}}\left[G_{10}\left(2-\lambda\right)+N_{s}G_{8}\right]d\lambda,
\end{align}
where $\omega_{A}=v_{A}/qR_{0}$ and $v_{A}=B_{0}/\sqrt{4\pi n_{i}m_{i}}$.

\bibliographystyle{iopart-num}
\bibliography{MyLibrary}

\end{document}